\documentclass[namedreferences]{kluwer}
\usepackage{graphicx}
\usepackage{alltt}
\usepackage{graphics}
\usepackage{epsfig}
\usepackage{color}

\begin{document}
\begin{article}

\begin{opening}

\title{Fast  object detection for use  onboard satellites}

\author{Martin \surname{Bange}\email{bange@faw.uni-ulm.de}\thanks{corresponding
         author}}
\institute{Forschungsinstitut f\"ur Anwendungsorientierte 
Wissensverarbeitung Ulm, Helmholtzstrasse 16, D89081 Ulm, Germany}
\author{Stefan \surname{Jordan}\email{jordan@astro.uni-tuebingen.de}}
\institute{Institut f\"ur Astronomie und Astrophysik, Universit\"at
T\"ubingen, Sand 1,
         D72076 T\"ubingen, Germany}
\author{Michael \surname{Biermann}\email{mbierman@lsw.uni-heidelberg.de}}
\institute{Landessternwarte Heidelberg, K\"onigstuhl, D69117 Heidelberg, 
Germany}
\author{Thomas \surname{K\"ampke}\email{kaempke@faw.uni-ulm.de}}
\institute{Forschungsinstitut f\"ur Anwendungsorientierte 
Wissensverarbeitung Ulm, Helmholtzstrasse 16, D89081 Ulm, Germany}
\author{Ralf-Dieter \surname{Scholz}\email{rdscholz@aip.de}}
\institute{Astrophysikalisches Institut Potsdam, An der Sternwarte 10,
D14482 Potsdam, Germany}


\runningtitle{Fast object detection}
\runningauthor{Bange et al.}

\begin{abstract}
We propose an object detection algorithm which is efficient and
fast enough to be used in (almost) real time with the limited 
computer capacities onboard satellites. 
 For stars below the saturation limit of the CCD detectors it is
based on a four neighbourhood local maximum criterion in order to
find the centre of a stellar image. For saturated stars it is based
on the assumption that the image is increasing monotonically towards
the centre in the unsaturated part of the image. The algorithm also
calculates approximate stellar magnitudes and efficiently rejects
most of the cosmics which would otherwise lead to a large number of
false detections.  The quality of the algorithm was evaluated with
the help of a large set of simulated data for the DIVA satellite
mission; different assumptions were made for the noise level, and the
presence of cosmics or for a variable sky background.
We could show that our algorithm fulfills the requirements for
DIVA; only in the case of simulated images which included 
the bright galaxy  M31 some fainter stars could not be detected in
the galaxy's vicinity. Since stellar images contain large areas without
any stars, we propose an additional block-skipping algorithm which can
be coded on special-purpose hardware. 

\end{abstract}
\keywords{CCD image simulation, noise modeling, real-time segmentation.}

\end{opening}


\sloppy

\section{Introduction}

Future all-sky space astrometric missions produce huge amounts of data that
cannot be sent down to earth. Therefore either only those parts of the sky
which are of interest for astronomers (namely small patches around stars
or other objects) have to be handpicked for telemetry or all astrometrically
relevant object parameters have to be computed already on board. The latter
will be performed on GAIA, the next ESA astrometric space mission, whereas
the former will be chosen for reduction of the downlink rate for DIVA.
An object detection algorithm, suitable for DIVA to extract such small patches
around detected objects, will be presented in this paper;
it is general enough to be adopted to similar mission
requirements on other satellites.

DIVA is a small astrometric satellite planned for launch in 2006. It will
provide high-precision positions, proper motions, parallaxes and photometric
data for some 35 million stars, and spectrophotometric information for about
12 million stars. 

\begin{figure}[t]
\begin{center}
\psfig{figure=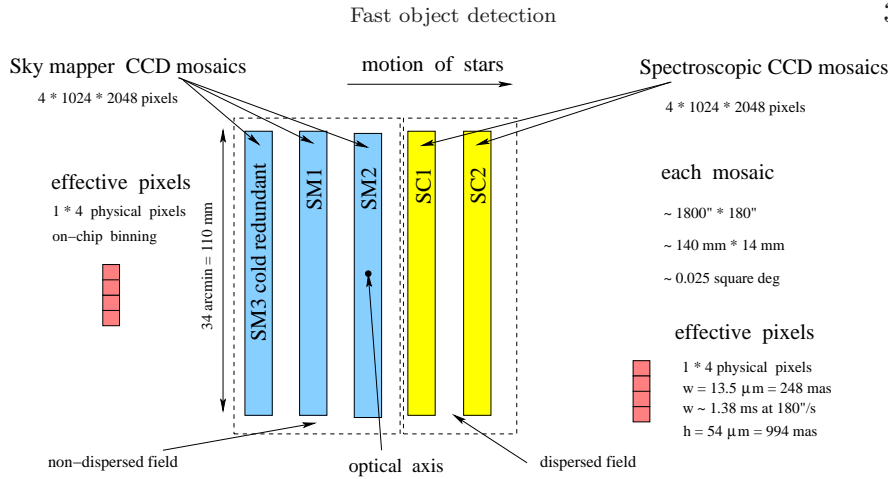,width=\textwidth,angle=0}
\caption{The focal plane layout of DIVA. Object detection will be performed
  on the four Sky Mapper CCD chips SM1 and in those regions on the slightly
  shifted SM2 CCDs, which were missed due to gaps between the SM1 chips.}
\label{fp}
\end{center}
\end{figure}

A major technical constraint of the DIVA satellite mission is the limited
data downlink rate. While 100 Mbit/s of raw data are produced by the DIVA 
detectors, only 0.7 Mbit/s can be sent to the ground for 17 hours each day.
No loss-less compression algorithm is able to account for such a large
discrepancy. Moreover, DIVA follows the conservative approach that
the astrometric data analysis is completely performed on ground.
This, however, means that all data relevant for the astrometric,
photometric, and spectroscopic measurements of DIVA have to be selected in real
time. For the Sky Mapper CCD detectors (SM1, SM2, see Fig.\,\ref{fp}) these are
the contents of small rectangular patches of  CCD pixels surrounding an object
(single
or multiple star).
Intensive studies with simulated data have shown
that sizes of $12\times 17$ are sufficient for stars with DIVA magnitudes
$D< 9.5$ and $12\times 7$ for stars down to the limit of
$D=15.2$. 
DIVA magnitudes are similar to $R$- and $I$-magnitudes and have their maximum
sensitivity at about 600 nm.

This article mainly deals with the detection and centroiding of stars.
Nevertheless, the algorithm presented here also detects non-resolved
double stars as well as the dense regions of cores of globular clusters
which cannot be resolved by DIVA or any other astrometric space mission
but which are well above any reasonable detection limit. In principle, this
algorithm can also detect minor planets in the solar system, but this was
not tested and is not subject of this paper.

Moreover, it is important to distinguish the objects of interest from detector 
artifacts, particularly cosmic ray hits (short: cosmics) and bad
columns. Both tasks -- object detection and artifact elimination -- have to be
performed for SM1 in almost real time on the IPU (Instrument Processing Unit)
with only a limited amount of buffer capacity; additionally, the positions
of such  objects have to be identified on the SM2 detector which were
missed  due to the gaps between the four
SM1 CCD chips.

The
scientific data of DIVA do not only consist of the rectangular patches on the SM
detectors.
Brighter stars ($D< 13.3$) are also seen as spectroscopicly dispersed
images (dispis) on the SC detectors (see Fig.\,\ref{fp}). Their positions are
predicted from
the centres of the objects found by the SM1 (and in a few cases on
SM2, respectively).
Rectangles containing the dispis  are cut out of the SC detectors for further
analyses on ground.

DIVA will capture image data in time delayed integration (TDI) mode where
the CCD read out is synchronised with the satellite rotation. As a
consequence, there are no images in the usual sense of frames but a
continuous stream of single lines.  For our simulation we will always use
1158 of such lines, corresponding to 1.5~seconds of observations.
The resulting frame size $1158 \times 525\times 4$ (4 CCDs) corresponds to
the amount of SM data that can be stored in the IPU ringbuffer.

Section\,\ref{problem} specifies the requirements that must be fulfilled by
any appropriate object detection algorithm. Section\,\ref{simulation}
presents the simulated data, which is the input of the proposed object
detection algorithm presented in Section\,\ref{algorithm}. The efficiency
and quality of the algorithm is described in Section\,\ref{evaluation}.

\section{Requirements}
\label{problem}
The object detection algorithm  should deliver at least 3 output parameters.
First of all, the centre of the detected object in pixel coordinates of the
corresponding CCD chip characterises the position of the object uniquely. Secondly,
the
magnitude of each object is needed, in order to decide if this object must be cut
out on the SC chips and to reject stars fainter than a given threshold.
Thirdly, cosmic-ray hits have to be distinguished from single stars,
double stars and extended objects like the cores of
globular clusters.

Around the predicted centre of a detected object some small rectangular patch
of pixels is cut out for telemetry. To make sure that all relevant information
is included in the pixel array the computed centre
 of the object must not
differ from the real centre of the object by more than 2 effective pixels in
each direction.

For on-board software there are additional constraints that do not hold for
on-ground software. Firstly, the available RAM for code and data is very limited.
It will amount to only 256 or 512 kByte for the  DIVA project.
Therefore it is not possible to process large parts of data. In fact an
effective object detection algorithm should work as local as possible 
in the sense of a  a very limited number of CCD
lines only. Secondly, the computational power of space-qualified processors is
still very low: DIVA will fly with 20 MHz processors so that  any code must be
designed optimised to meet these requirements.

The quality of an  detection algorithm is judged by comparison with the 
detection probabilities that are predescribed for a particular satellite
mission \cite{BB}. For DIVA  they are 
presented in Table \ref{probability}. 

\begin{table}[H]
\begin{tabular}{cc}
\hline
magnitude                   & detection probability \\ 
\hline
saturated stars ($D<8$)     & $\ge 95$ \% \\
$8\le D < 14.6$                  & $\ge 99.5$ \%  \\
$D = 15.2$                  & $(50\pm 5)$ \%  \\
\hline
\end{tabular}
\caption{Required detection probability as a function of the object magnitude}
\label{probability}
\end{table}

\section{Simulation}
\label{simulation}

A sufficiently large amount of simulated SM data including several
thousand stars as well as realistic noise, sky and CCD properties is
needed in order to investigate the on-board data reduction.
The efficiency of the object detection algorithms was tested with
128 simulated frames for the  different SM CCD chips.
The data have been simulated \cite{Scholz} based on real
sky observations and star catalogues with a preference given to
relatively high stellar density fields, also including some open clusters.
 
\begin{figure}
\begin{center}
\psfig{figure=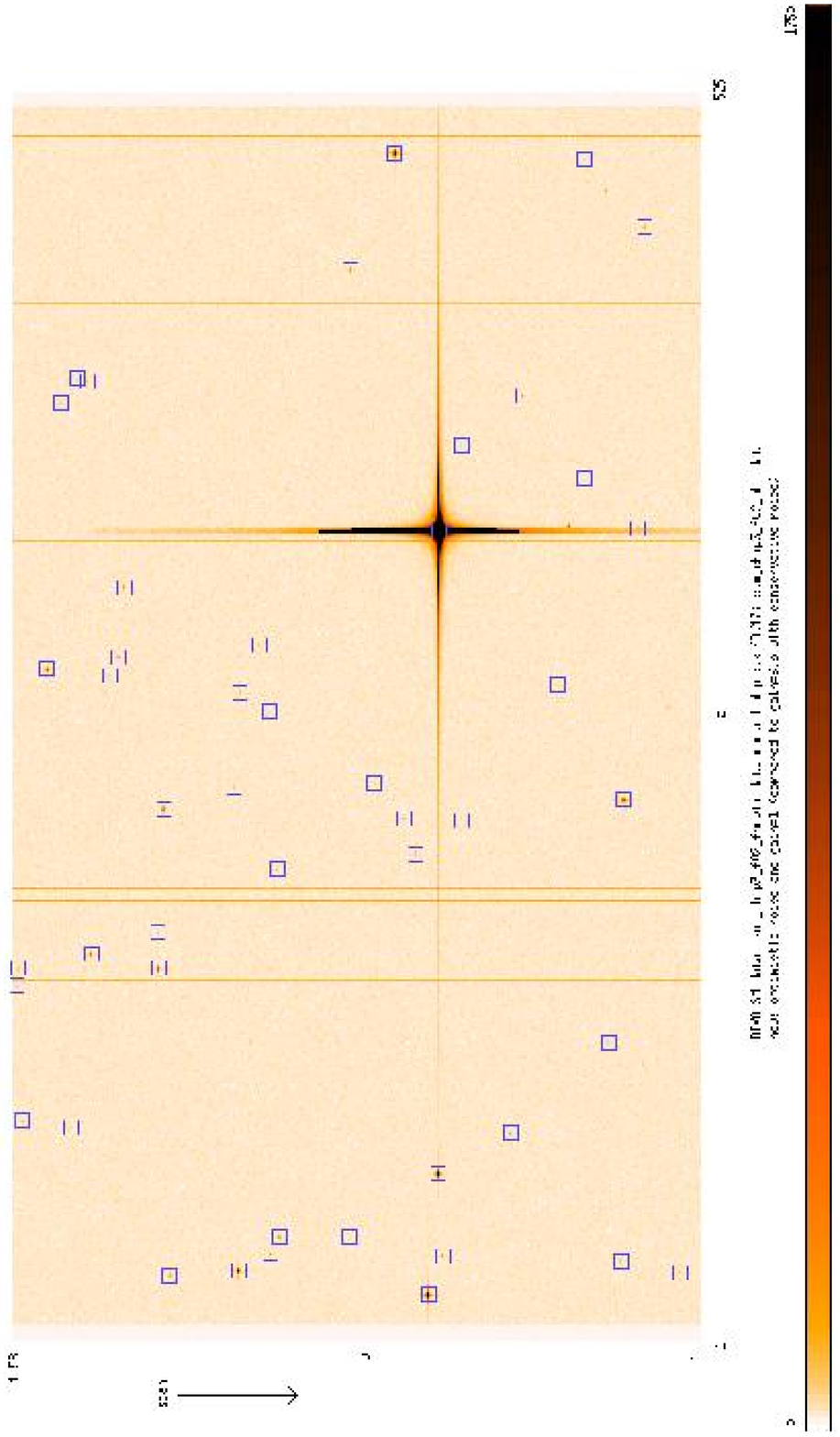,width=\textwidth,angle=0,clip=}
\caption{SM chip field of normal stellar density with optimistic noise and
without cosmics. Only the brighter simulated objects ($D<17$) are marked.
A very bright saturated star ($D=2.5$) with its large diffraction spikes has
been included in the simulations.}
\label{fig_smchip1}
\end{center}
\end{figure}
 
\begin{figure}
\begin{center}
\psfig{figure=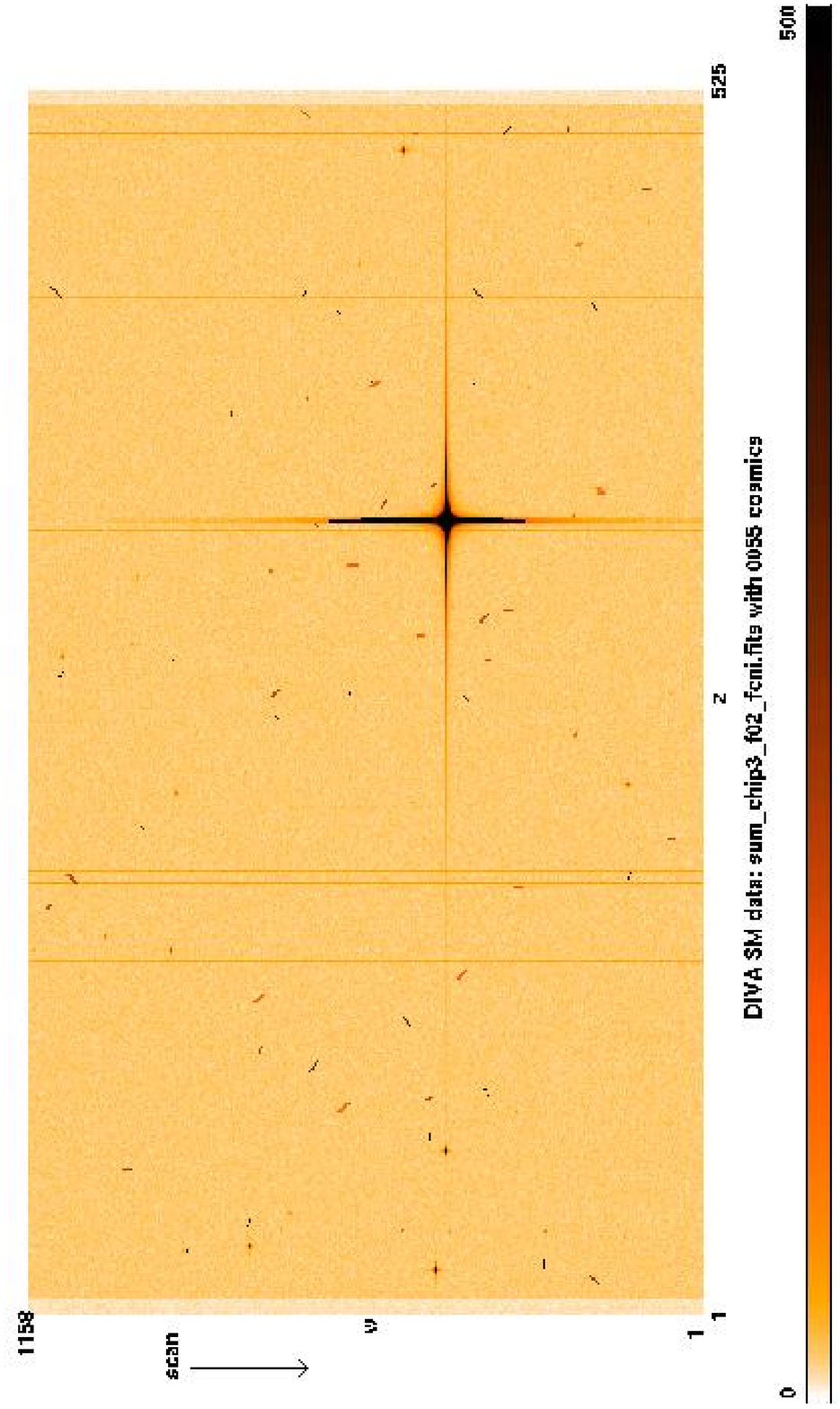,width=\textwidth,angle=0,clip=}
\caption{The same SM chip field with conservative noise and with cosmics.}
\label{fig_smchip2}
\end{center}
\end{figure}
 
Additionally, 4 of the frames contained the image of the globular
cluster M4 (which is relatively sparse). On two frames the dense globular
cluster M30 was included in the corner.  All stars up to
the faintest ones were simulated in case of M30, for which Hubble Space
Telescope  observations
\cite{GuWeYaScBa} served as input data. On one eighth of all frames  a
large number of faint (background) objects were present, representing the
situation when DIVA scans the Galactic plane; for this purpose we used
a catalogue of stars in Baade's window
\cite{SzUdKuKaMaKr} as input data.
 
Each CCD chip was modeled with its own  characteristic hot and dark
columns and a slightly different  sensitivity function ($>90\%$) over
 the CCD columns
(one-dimensional flat field). Conservative noise parameters (dark current
of 5 electrons per original pixel and per second, read-out-noise with
$\sigma$ of 7 electrons per effective pixel) and a gain of 3.5 were used,
i.e. one signal unit corresponded to 3.5 electrons. A bias level of 
30 electrons per effective pixel was added. 
 
The whole data set has also been simulated with optimistic noise
parameters (see Fig.~\ref{fig_smchip1}). The optimistic and conservative
CCD noise parameters are defined in the DIVA mission requirements,
respectively as optimum and minimum requirements. 
Optimistic parameters were defined by a dark current of 2 electrons per
original pixel and per second and a read-out-noise with a $\sigma$ of 2 electrons
per effective pixel. In case of optimistic noise, a gain of 1 was used,
i.e. one signal unit corresponded to 1 electron.
 
However, the basic simulated data set used in this investigation was that 
with conservative noise, and it did also include cosmics 
(see Fig.~\ref{fig_smchip2}) with a mean number of 70 events
per chip field. Cosmics were simply simulated as connected pixels of
constant signal (typically 500 electrons). The length of the cosmic ray
traces may have been overestimated. The above numbers were estimated for
GAIA from experience with the Hubble Space Telescope CCD chips  \cite{ESA2000}. 

All simulations included many  faint stars far below the envisaged
detection limit of $D\sim15.2$. In total we have considered 19000 stars
with 7000 being brighter than $D=16.1$.
Saturated bright stars have also been
simulated and added to the data stream with a simplified model on the
transfer of charges above the saturation level: Effective pixels with
a signal of $>150000$ electrons were considered saturated and the charges
above the saturation were distributed along the CCD columns in the scan
direction (and opposite direction) with a slight asymmetry (more charges
transferred in opposite scan direction). A maximum of one saturated star with
($2 \le D \le 8$) were randomly added to 27\%\ of the   files
which is much more frequent than the average during the
DIVA mission (about 2\%).

In order to be able to test our algorithm for the case of the strongest
variation of the background during the DIVA measurements, 
the M31 galaxy was simulated 
in the centre of the chip fields and
with its major axis in cross-scan direction. The maximum signal in
the core of M31 was estimated as 410 electrons per effective pixel.
Near to the edge of the chip field (which is about $290 \times 525$
arcsec$^2$),
the signal from M31 is only about 5 electrons per effective pixel.
The conclusion is, that there are no serious background variations
to be expected due to extended objects. M31 is certainly one of the
brightest extended objects on the sky, but with the short exposure
time of 1.5 seconds and the relatively small aperture of DIVA, we
see only the core region of M31.

\section{Algorithm}
\label{algorithm}

In computer vision the detection of stars and their positions with
CCD detectors would be called
a segmentation problem. In  principle,
segmentation algorithms for star images may be designed from  
two extreme perspectives, that of  ''growing'' and that of
''shrinking''. In the first case an
image spot is tentatively or permanently considered to be the centre
of a star; then a region is grown or flooded to eventually contain the star.
In the other extreme, a candidate region of the image is selected which
is likely to contain at least one star. Actual star presence is then
tested for the candidate region. In the affirmative case, all star
centres must be localised. In both cases, star regions may overlap. Due
to conceptual and computational simplicity, the first perspective is
adopted here.

In any case, fast segmentation algorithms are
typically based on local operations. In astronomy, known examples of
such algorithms are SWA (sliding window algorithm) \cite{SWA} and
SExtractor (source extractor) \cite{Be}. In computer vision in general,
object detection under the given time constraints is related to real-time
image segmentation for video processing, see for example~\cite{LiKeHjPi}.
The fundamental difficulty with local segmentation methods is their weak
adaptation to global features, such as  a slowly varying background level.
However, computational studies of adaptive image segmentation algorithms
based on entropy \cite{Ot} and conditional entropy \cite{JoBi}
revealed unacceptably high computation times for the present problem.
To what extend a  varying background may be a problem  
is considered by testing our algorithm on frames containing the bright
galaxy M31.

We propose a simple detection and centroiding algorithm for
DIVA which is motivated by the fact that a local maximum in a stellar
image is very close to the position (centre) of the star. However, for
stars above the saturation threshold of the CCD detector a different
treatment is necessary.

The proposed algorithm identifies pixels that are local intensity maxima
and that exceed a certain threshold. These pixels become candidates for
star centres. Each candidate undergoes several tests and is accepted as
star centre if and only if it passes all of them. All tests are local
(and therefore fast) 
considering at most a $3 \times 3$ window around the centre candidate
pixel.

More precisely, a pixel $p$ is selected as the centre of a stellar
candidate if the pixel intensity $I(p)$ exceeds the detection threshold
\begin{equation}
T = B + 4 \cdot \sqrt{B},
\end{equation}
where $B$ is the background. Here the assumption was made that the noise
is poissonian.
This is no serious limitation because the factor of
4 can be easily changed to a different value; slight adaptations are always
possible during the commissioning phase of DIVA or whenever necessary.
For the background value $B$ we used the median of the simulated images since
it is a very robust estimate in almost all astronomical images.
For the on-board IPU software of DIVA we plan to use an approximate median
from 81  points for several sampling regions on the SM detectors. The value
is calculated by a special purpose hardware in a 4 stage pipeline, which in
each step calculates the median of triples only (see \cite{DIVA}). In this
paper we assume a constant background over one simulated frame. For DIVA,
however, we plan smaller sampling regions to better account for 
varying background.

Since CCD detectors have a limited dynamic range, a special treatment is
necessary for saturated stellar images. Therefore, we have to distinguish
between star candidates above saturation threshold $T_S$ and those below this
limit but exceeding the detection threshold.

The pseudo code of this procedure is as follows.

\begin{alltt}
/* Main part */
T = bg + 4 * squareroot(bg); /* bg = median of the image */
FOR (every image row) DO
     FOR (every pixel in the current row) DO
       IF (intensity > \(T\)) AND (intensity \(\le T\sb{S}\)) THEN
         analyse_unsaturated_object();
       ELSEIF (intensity \(> T\sb{S}\)) THEN
         analyse_saturated_object();
       ENDIF
     ENDFOR
     IF (analysis of a saturated object is completed) THEN
       save object coordinates;
     ENDIF
ENDFOR
\end{alltt}

\subsection{Unsaturated stars}

The analysis for an unsaturated star is based on the intensity $I(p)$ of
the current pixel $p$ exceeding the detection threshold $T$ and staying
below the saturation limit $T_S$. In order to be the centre of a star the pixel
$p$ has to pass some additional tests.

\begin{figure}
\begin{center}
\psfig{figure=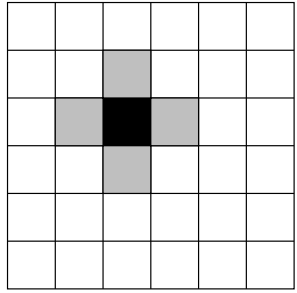}
\caption{The currently considered pixel (black) with its neighbourhood of four
pixels which are tested for having a smaller value.}
\label{fig_neighbour}
\end{center}
\end{figure}

One feature of an unsaturated star is that its centre represents a local
intensity maximum. Therefore $I(p)$ has to exceed the value of each of its
neighbouring pixels (neighbourhood of four, see Fig.\,\ref{fig_neighbour}) or
at least the value of three
neighbouring pixels and to  be equal to the fourth one. The latter case can
be caused by noise. It is possible that even more than two
pixels in the centre area of an unsaturated star have the same value. But
in practice this is unlikely.

In order to be regarded as a star, $N$ of the four neighbours
have to exceed the detection threshold. $N$ was determined experimentally
and depends on the data set. For conservative noise $N = 2$ was chosen
and for optimistic noise $N = 3$ was chosen. The value had to be lowered
for conservative noise because otherwise too many stars in the magnitude
range $8.0 \le D < 14.6$ would have been lost. This condition helps to
eliminate some cosmics, noise peaks and detections on hot columns.

If the current pixel passed the foregoing tests, measures for the star
magnitude and a measure for the likelihood to be a cosmic are computed. In
order to determine  
the magnitude, the $3 \times 3$ window $W$ surrounding the current
pixel is used. Only those pixels $p_i$ in this window with an intensity
$I(p_i)>T$ are taken into account for the magnitude computation which is
as follows:
\begin{equation}
m = \sum_{\{p_i \epsilon W | I(p_i) > T\}} (I(p_i) - B)
\end{equation}
This value must exceed a magnitude threshold $T_m$. Otherwise the current
pixel cannot belong to a sufficiently bright star. $T_m$ depends on the
data set and will be calibrated during the commissioning phase.
Simulations have shown that the accuracy of the magnitude determination
($1\sigma$) is $\Delta D\approx 0.15$ at $D<15$ and $0.5$  at $D \ge
15$. For saturated stars no magnitude determination was performed.
A serious problem arises from cosmics. As described in Section
\ref{simulation} we can expect about 70 cosmics per image. Many of them
will pass the three previous tests. Moreover, the algorithm often detects
more than one local maximum per cosmic (in the average), increasing
the rate of false detections extremely. Hence it is crucial that the
algorithm is able to filter out most of such false detections. In
contrast to a star, the intensity values of a cosmic form a plateau with
arbitrary orientation. This is different from the appearance of a star
where we have a sharp maximum which is significantly higher than the
average intensity of the other pixels belonging to the star image.
Thus, for cosmic rejection, 
the difference
\begin{equation}
I(p) - \frac{m}{n}
\end{equation}
can be used as indicator, where $n = |\{p_i \in W | I(p_i) > T\}|$.
This difference is sufficient to distinguish between cosmics and bright
stars. However, for fainter stars close to the DIVA limit of about
$D=15.2$ we had to take the dependence of this difference from the
magnitude of a star into account. Therefore, the above difference is set
in relation to the average magnitude per pixel. Cosmics have a higher
average magnitude compared to faint stars. The fraction 
\begin{equation}
c = \frac{I(p) - m/n}{m/n}
\end{equation}
is used to distinguish stars from cosmics and is a measure
for the likelihood to be a  cosmic. In order to treat an object
as a cosmic, $c$ 
must exceed a threshold $T_c$ which depends on the data
set and which has to be calibrated during the commissioning phase.
This criterion also helps to sort out some false detections on hot
columns. The pseudo code of this analysis is as follows:

\begin{alltt}
/* analyse_unsaturated_object() */
IF (current pixel is a local maximum) AND
   (N of the 4 neighbours exceed T) THEN
     IF (magnitude \(> T\sb{m}\)) AND (\(c > T\sb{c}\)) THEN
       save object coordinates and magnitude value;
     ENDIF
ENDIF
\end{alltt}

\subsection{Saturated stars}

A saturated star image, which contains at least one pixel above the saturation
threshold $T_S\gg T$, can differ drastically from that of unsaturated stars
so that a special treatment is necessary. In this case a mere local
maximum search over intensities does not find the centres of
saturated stars within the required precision. For this reason, the
detection algorithm for saturated stars allows for a mixed mode of
maximum search. This mixed mode considers intensity values as well as
sets of adjacent pixels above saturation threshold. 

The treatment of saturated star images depends on the point spread function
in the focal plane. Our algorithm is based on the assumption that
the point spread function is monotonically increasing  towards the centre of
the star, which is the case of the cross-like star image caused by
the rectangular apertures of the DIVA telescopes.

Let $(i,j)$ be a pixel above saturation level. Let this pixel
be the first saturated pixel in the scanning sequence which means that its
left neighbour (in same row $i$) and
its upper neighbour (in same column $j$) are unsaturated.

Scanning the preceeding row $i-1$ leads to the column
coordinate of the centre of the saturated star.
Starting from pixel $(i-1, j)$, row $i-1$ is scanned to the left as long as
intensities increase in this direction. Then the maximum search is continued
from pixel $(i-1, j)$ in the right direction, again as long as intensities
increase. The column index of the resulting maximum corresponds to the column
index of the centre.

The row index of the centre is found as follows: Beginning with pixel
$(i,j)$, a set of consecutive pixels in row $i$ is scanned to the right
until intensities drop below the saturation threshold. This is repeated
for saturated pixels $(i+1,j)$, $(i+2,j)$ etc. as long as the number of
these adjacent pixels above saturation threshold increases from row to
row. The scanning extends to both the left and the right direction for
rows $i+1$, $i+2$ etc. The row of the largest of such intervals is
selected as the row index of the star centre. This contains a
monotonicity criterion: The number of connected saturated pixels
increases from row to row as long as the centre of the star is not
reached.

\begin{figure}
\begin{center}
\psfig{figure=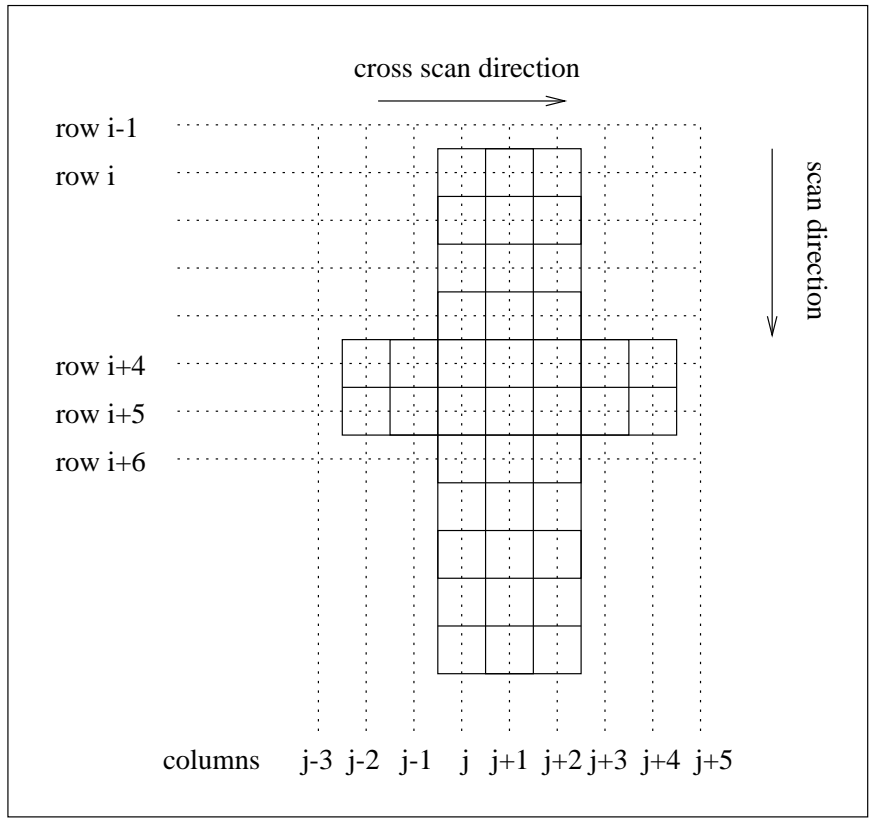}
\caption{Illustration of the algorithm for saturated stars.}
\label{fig_satdetect}
\end{center}
\end{figure}

Fig.\,\ref{fig_satdetect} shows the saturated pixels of a star and illustrates
the algorithm. Technically, the search for the row index is done only
towards the right.  It may then happen, that one
saturated star is scanned starting from different saturated pixels.
Different candidate regions of saturated stars are treated as one single
object if they have one saturated pixel in common.

\begin{alltt}
/* analyse_saturated_object() */
determine number S of connected saturated pixels of the
    star in the current row;
IF (current row is the first row that contains saturated
    pixels of the current object) THEN
    search the pixels in the previous row that belong to the
      current object for their maximum;
    cross scan coordinate of centre = cross scan coordinate
      of the found maximum;
    /* Initialise scan coordinate of the centre */
    scan coordinate of centre = index of the current row;
    maximum in scan direction = intensity of the first
      unsaturated pixel to the left or to the right of the
      saturated pixels;
ELSE
    IF (S(current row) > S(previous row)) THEN
       scan coordinate of centre = index of current row;
       maximum in scan direction = intensity of the first
         unsaturated pixel to the left or to the right of
         the saturated pixels;
    ELSEIF S(current row) == S(previous row) THEN
       IF (maximum in scan direction) <= (intensity of
         the first unsaturated pixel to the left or to the
         right of the saturated pixels) THEN
         scan coordinate of centre = index of current row;
         maximum in scan direction = intensity of the first
           unsaturated pixel to the left or to the right of
           the saturated pixels;
       ENDIF
    ENDIF
ENDIF
\end{alltt}

\subsection{Acceleration by block skipping}

The star detection algorithm  as developed so far compares every 
pixel value with
the detection threshold. But large parts of the sky are empty in
the sense that there are no interesting objects, i.e. no pixels with
intensities above the threshold. If the algorithm can skip these parts,
computation time will be saved.

Therefore, the first $k$ pixel values of a row are compared to the
detection threshold $T$. If one of these intensities exceeds the
threshold, a header bit for this block is set to one. Otherwise the header
bit is set to zero. Then the next $k$ pixel values of this row are
compared to $T$. Accordingly to the result a second header bit is set.
This is continued until the end of the row. This procedure is executed
for each row. The star detection algorithm will be run only for those
blocks whose header bits equal one. It is run for all pixels of such
blocks. Such a simple identification of interesting blocks can be easily
done with special purpose hardware chips and would therefore relieve the
on-board computer significantly.

\section{Evaluation}
\label{evaluation}
The algorithm was tested with all four sets of simulated data
(conservative and optimistic noise, each for the case with and
without cosmics).
The results for the different data sets are shown in the 
Table\,\ref{tab_con_noise}. 
The first column in our tables of results
describes the real (predescribed) magnitude of the star, not the computed one.
In the following columns the detection rates for all data sets are given.
A star is regarded as detected correctly if the centre determined by the
algorithm agrees with the simulated image location within two pixels in both
coordinates and if the detected star is brighter than $D = 16.1$.
On the other side a false detection occurs if no simulated star image brighter
than $D = 16.1$ lies within two pixels of the detection's location indicator
in both coordinates.

\begin{table}[H]
\caption{Detection rates for the data sets with conservative noise
and optimistic noise, each  
including and without  cosmics. The corrected false detection
rate takes into account that the number of false detections which are
due to bright stars is lower in reality by a factor of 14. 
}
\label{tab_con_noise}
\begin{tabular}{crr}
\hline
Magnitude                   & det. rate w.o. cosmics & with cosmics \\  
& conservative noise & conservative noise  \\  \hline
$D < 8.0$ (saturated)       & $100\%$ & $100\%$   \\
$8.0 \leq D < 14.6$         & $99.6\%$ & $99.5\%$ \\
$D \approx 15.2$            & $47.1\%$ & $47.5\%$ \\
False detections            & $16.8\%$ & $21.3\%$  \\
False detections (corr.)    & $2.1\%$ & $6.0\%$  \\
\hline
Magnitude                   & det. rate w.o. cosmics & with cosmics \\  
& optimistic noise & optimistic noise  \\  \hline
$D < 8.0$ (saturated)       & $100\%$  &$100\%$   \\
$8.0 \leq D < 14.6$         & $99.7\%$  &$99.6\%$ \\
$D \approx 15.2$            & $51.5\%$  &$49.0\%$ \\
False detections            & $12.9\%$  & $16.2\%$  \\
False detections (corr.)    & $1.8\%$  & $6.2\%$  \\
\hline
\end{tabular}
\end{table}

It turned out that the algorithm is successful in  centroiding 
saturated stars. For $8.0 \leq D < 14.6$ it also fulfills the requirement
of 99.5\%, specified for the DIVA satellite mission. Two of the simulated
stars in this magnitude range (corresponding to about $0.1$\%) could not be
detected because they were too close to a much brighter star. Therefore,
the centres of the two fainter stars do not represent local maxima in this
environment. 

The cosmic criterion allows us to achieve the required detection rates
for $D \approx 15.2$ while keeping the false detections in an acceptable
range. 
Most of these false detections are caused by local maxima
that arise on the axes of saturated stars because of noise. This problem could
be further mitigated with the aid of the cosmic criterion but only at the
expense of the positive detections ($8.0 < D \le 14.6$ and $D \approx 15.2$).
A minor part of the false detections is caused by hot columns; they were
reduced
considerably by a division of the bad columns by a factor of 2. This factor
has to be optimized during the commissioning phase.
In reality the number of false detections would be much smaller, since
the majority of them are due to the presence of bright saturated stars.
Since our simulations have included a  factor of 14  more bright objects
than in an average region of the sky, we have listed a corrected value in
the table for which we have divided the number of false detections which were
due to bright stars by this factor.

\subsection{Results for M31}
In order to test whether a variable background would  affect
the detection rate of our algorithm we additionally applied it to 
the simulated data sets containing the galaxy  M31 
(see section \ref{simulation}).
M31 can be considered as a worst case since it has a very large surface
brightness in its core.  
For these data sets we obtained the following results:

\begin{table}[H]
\caption{Detection rates for the data set containing M31 assuming
conservative noise and the presence of cosmics. 
Since most of the false detections were due to the bright core of M31
a corrected false detection rate was not listed.
}
\label{tab_M31}
\begin{tabular}{cc}
\hline
Magnitude                   & detection rate  \\      \hline
$D < 8.0$ (saturated)       & $100\%$   \\
$8.0 \leq D < 14.6$         & $99.6\%$  \\
$D \approx 15.2$            & $39.7\%$  \\
False detections            & $511.8\%$   \\
\hline
\end{tabular}
\end{table}

The presence of M31 has almost no influence on the rate of successful
detection. The only problem is caused by false detections in the very
neighbourhood of the galaxie's core. This is very similar to some false
detections caused by noise peaks close to bright saturated stars.
M31's core creates an area on the CCD which is above the detection threshold
so that this single criterion is always fulfilled. If then, due to noise, 
a pixel is a local maximum, it is treated as a star.
However, even the high number of false detections would cause no real
limitations, because they are restricted to extremely small areas of
the sky. Although the described algorithm has not been tested on the
DIVA computer hardware, we believe that this peak can still be handled
(from conservative estimations we expect that we can handle about 300-1000
detections per second). Since M31 can be regarded as a worst case, we do
not expect similar problems elsewhere. If we were, due a data overflow, forced
to disregard a part of the stars in this region, this would only lead
to a smaller number of measurements since not always the same stars would
be lost. 

\section{Conclusion}
Our very simple four-neighbourhood local maximum method  together with
some special algorithms for saturated stars, cosmic detection, and magnitude
determination
turned out to
fulfill the requirements of astrometric satellite missions like DIVA: It
is fast, finds the centres of nearly all stars down to $D=14.6$ and
most at the limit of 
$D=15.2$, rejects cosmics very efficiently, calculates the
stellar magnitudes with sufficient precision, and has a very low number of
false detections. The number of wrong identifications only increases in
the neighbourhood of bright extended objects like extremely bright and saturated
stars or the cores of bright galaxies. Since such peaks  correspond to only very
small areas of the sky (corresponding to only fractions of a second during
DIVA measurements) they do not pose major difficulties.

A comparison of our method with other local segmentation algorithms has
shown it's superiority in both speed and reliability. 
As an example we have tested the efficiency of the local
SExtractor
 and SWA algorithms  and found that 
their ability to detect bright and saturated stars was well below the
requirements. 
Alternatives to local detection like adaptive image segmentation algorithms
based on entropy  and conditional entropy would cause a need for 
more memory for the on-board computer than usually available on space
missions.

\acknowledgements
Algorithm development was supported in part by Astrium company under DLR
grant 50\,QD\,0106/F\,39260. 
Stefan Jordan and Michael Biermann were  supported by  DLR 
 grants 50\,QD\,0109 and 
50\,QD\,0101, respectively.

\end{article}
\end{document}